\documentclass[aps,preprint,prb,amssymb,12pt,floatfix]{revtex4}
\usepackage{amsmath}
\usepackage{subfigure}
\usepackage{graphicx}
\usepackage{lscape,graphicx}
\usepackage{rotating}
\usepackage{epstopdf}
\usepackage{color}
\usepackage{amsmath,amsthm}
\usepackage{wrapfig}

\def\la{\langle}
\def\ra{\rangle}

\def\fh{\mathcal{F}_h}

\begin{document}
\begin{center}
{\Large\bf On the origin of the unusual behavior in the stretching of single-stranded DNA}\\
\ \\
{\large Ngo Minh Toan$^1$ and D. Thirumalai$^{1,2}$}\\
\ \\
$^1$  Biophysics Program, Institute for Physical Science and Technology,\\
$^{1,2}$  Department of Chemistry and Biochemistry, University of Maryland at College Park, College Park, Maryland, USA, 20742
\end{center}
\date{\today}


\begin{abstract}
Force extension curves (FECs), which quantify the response of a  variety of biomolecules subject to mechanical force ($f$), are often quantitatively fit using worm-like chain (WLC) or freely-jointed chain (FJC) models. These models predict that the chain extension, $x$, normalized by the contour length increases linearly at small $f$ and at high forces scale as $x \sim (1 - f^{-\alpha})$ where $\alpha$= 0.5 for WLC and unity for FJC. In contrast, experiments on ssDNA show that over a range of $f$ and ionic concentration, $x$ scales as $x\sim\ln f$, which cannot be explained using WLC or FJC models. Using theory and simulations we show that this unusual behavior in FEC in ssDNA is due to sequence-independent polyelectrolyte effects. We show that the $x\sim \ln f$  arises because in the absence of force the tangent correlation function,  quantifying chain persistence, decays algebraically on length scales on the order of the Debye length. Our theory, which is most appropriate for monovalent salts, quantitatively fits the experimental data and further predicts that such a regime is not discernible in double stranded DNA.
\end{abstract}
\maketitle
\newpage

{\bf Introduction}:

The response of double stranded DNA, RNA, proteins and polysaccharides to mechanical force ($f$) has provided a microscopic basis for describing their elasticity~\cite{busta,WLC5,schulten3,gaub97,linke,marsza2,marsza1}. A number of studies have established that the measured force-extension curves (FECs) can be nearly quantitatively reproduced using standard worm-like chain (WLC) model~\cite{WLC1} or freely-jointed chain (FJC) model~\cite{DoiEdwards}. However, recent single molecule experiments on single stranded DNA (ssDNA) showed that the measured FECs exhibit unexpected behavior that cannot be described using WLC or FJC model. Surprisingly, over a wide range of stretching forces $f$ and at low ionic strengths $I$, the chain extension normalized by the contour length of the chain, $L$, scales as $x\sim \ln f$~\cite{dessinges,Koen,ssDNAPincus}.

The unusual behavior found for a range of $f$ values in ssDNA implies that the standard polymer models that take only the elasticity of the chain are not adequate. Here, we provide a theoretical explanation of the $x\sim \ln f$ behavior using analytical calculations and simulations by treating ssDNA as a polyelectrolyte chain. We show that the tangent correlation function from which the chain persistence length is extracted decays as a power law on scales $a\lesssim s \lesssim \lambda_D$ (the Debye length). Such a behavior, which was not previously noticed, is a consequence of polyelectrolyte (PE) effects in ssDNA and gives rise to the $x\sim\ln f$ behavior. We confirm the theoretical predictions using simulations, which also show excellent agreement with experimental data for ssDNA\cite{ssDNAPincus} for $I$ from 1mM to 3M. Our theory, not only produces quantitative agreement with experiments, but also establishes the importance of PE effects in predicting the response of ssDNA to $f$.

{\bf Theory for stretching ssDNA including polyelectrolyte effects.}

\begin{figure}
   \centering
   \includegraphics[width = 6in]{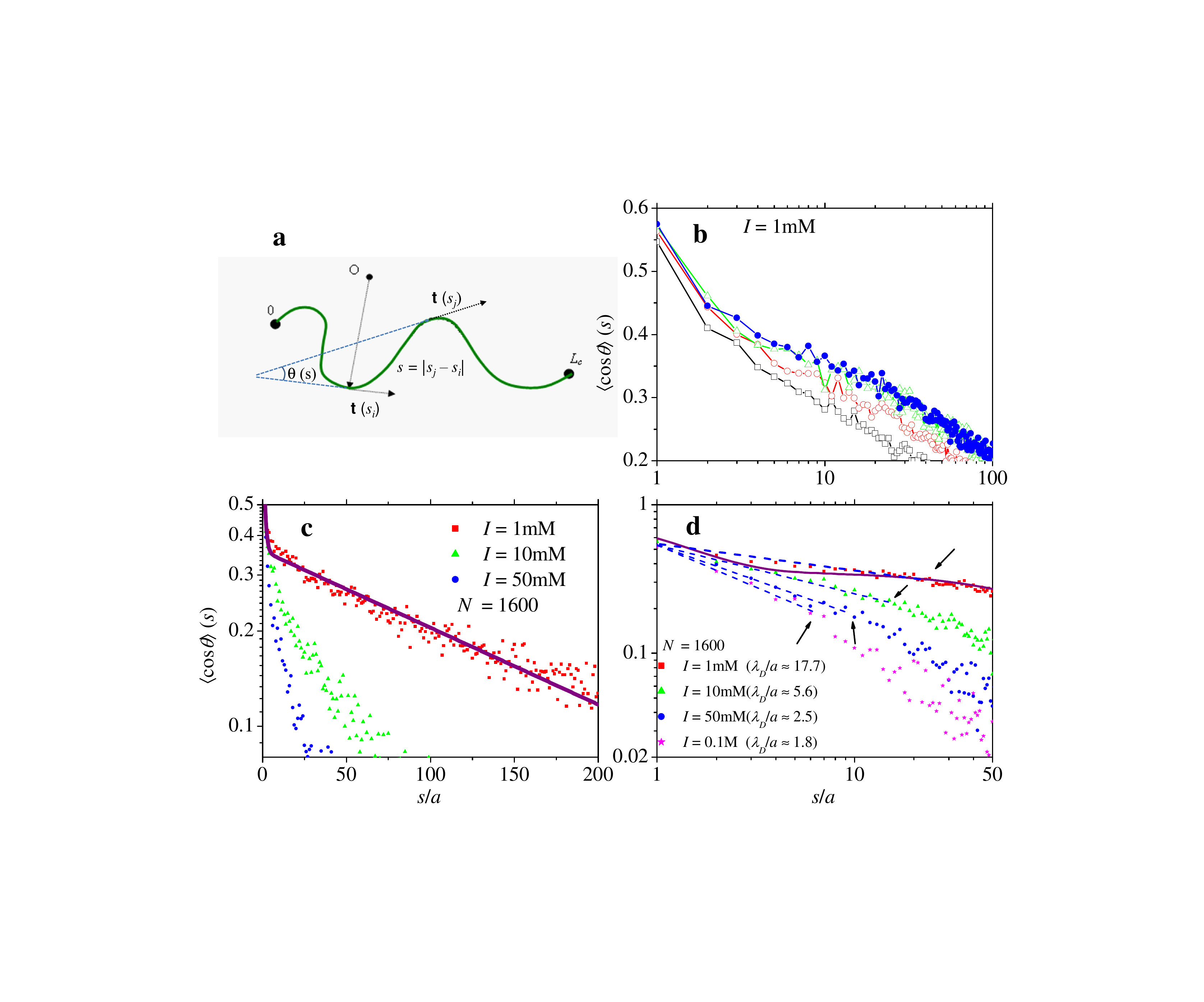}
   \caption{}\label{fig:fig1}
\end{figure}

Measured FECs for a broad class of biopolymers can be understood in terms of an interplay of the tensile length $\xi_t = k_BT/f$ ($k_BT$ is the thermal energy), the persistence length $\xi_p$ and the monomer length $a$. The resulting theory allows us to determine {\em a priori} the precise polymer model that best describes the measured FECs for a specific macromolecule~\cite{ToanMacro2010}. For a discrete semiflexible chain, the tangent correlation  at $f=0$ decays as
\begin{eqnarray}\label{eq:intrinsic_correlation_ds}
\la \cos\theta(s)\ra = \exp\left(-{s\over \xi_p}\right)
\end{eqnarray}
where $\theta(s)$ is the angle between two tangent vectors that are separated by length $s$ along the chain (Fig.~\ref{fig:fig1}a). At very high forces, $f > \fh \equiv ck_BT\xi_p/a^2$, $\la \cos\theta\ra$ is independent of $s$ and can be approximated as~\cite{ToanMacro2010}
\begin{eqnarray}\label{eq:force_induced_correlation_a}
\la \cos\theta\ra(f) \approx \exp\left(-c{\xi_t\over a}\right),
\end{eqnarray}
where $c$ is a model-dependent constant on the order of 4 for non-charged polymers. Thus, the macromolecule behaves as a FJC with $x \approx 1 - \xi_t/a \sim 1 - f^{-1}$. When $k_BT/\xi_p\le f\le\fh$, due to the interplay of $\xi_p$ and $\xi_t$, the macromolecule can be modeled by a sequence of independent segments whose length $\lambda_K$ is force-dependent. In this force regime, the tangent correlation is given by~\cite{ToanMacro2010}
\begin{eqnarray}\label{eq:force_induced_correlation_lk}
\la \cos\theta\ra(f) \approx \exp\left(-c{\xi_t\over \lambda_K}\right).
\end{eqnarray}
Thus, $\lambda_K = \sqrt{c\xi_t\xi_p}$ and $x \approx 1 - \xi_t/\lambda_K \sim 1 - f^{-1/2}$, which is the well-known WLC behavior\cite{WLC1,BYHa1997}.

Exponential decay of $\la cos(\theta(s)\ra$ in uncharged polymers (Eq.~\eqref{eq:intrinsic_correlation_ds}) results from the sequential transmission of interactions along the chain backbone .  However, in polyelectrolytes such as ssDNA, direct repulsive interactions between any pair of monomers within a scale $\sim\lambda_D$ could alter Eq.~\eqref{eq:intrinsic_correlation_ds}. It has already been shown that deviations from Eq.~\eqref{eq:intrinsic_correlation_ds} could occur even in stiff chains without intra molecular interactions~\cite{Hsu10Macro}. The PE effects can profoundly change the nature of FECs. Because the anticipated decay of $\la \cos\theta\ra$ in PEs, which is slower than in uncharged polymers affects FEC, we propose that the intrinsic ($f=0$) tangent correlation on scales $s \lesssim \lambda_D$ in polyelectrolytes should decay as
\begin{eqnarray}\label{eq:intrinsic_correlation_PE}
\la \cos\theta(s) \ra \approx C \left({s\over a}\right)^{-\gamma}, \text{for } a\le s \lesssim \lambda_D,
\end{eqnarray}
were $C$ and $\gamma$ are functions of $a/\lambda_D \propto I^{1/2}$. Note that, in the complete screening limit $\lambda_D/a \ll 1$, the correlation equation must reduce to the exponential form, thus $\la \cos\theta(s=a) \ra \equiv C_0 = \exp\left(-a/\xi_p^0\right)$, with $\xi_p^0$ being the bare persistence length; $C_0$ is, thus, the lower limit for $C$.

Just as for macromolecules with short-ranged interactions between the monomers, the FEC for PEs should quantitatively match the prediction of the FJC model as long as $f> \fh^{PE}$. We determine $\fh^{PE}$ by equating Eqs.~\eqref{eq:force_induced_correlation_a} and~\eqref{eq:intrinsic_correlation_PE} at $s = a$ yielding
\begin{eqnarray}\label{eq:fhPE}
\fh^{PE} = c {k_BT\over a \ln C^{-1}}.
\end{eqnarray}
In the complete screening limit $\fh^{PE}$ in Eq.~\eqref{eq:fhPE} reduces to $\fh$ for uncharged polymers described above.


In the regime where the force is high enough such that every segment on the order of $\lambda_D$ is independent of each other, the chain can be modeled as a FJC with force-dependent $\lambda_K$ segments, with the tangent correlation due to the tensile energy being given in Eq.~\eqref{eq:force_induced_correlation_lk}. By equating the {\em rhs's} of Eqs.~\eqref{eq:intrinsic_correlation_PE} and \eqref{eq:force_induced_correlation_lk} at $s=\lambda_K$ we have the solution for $\lambda_K(f)$,
\begin{eqnarray}
    \lambda_K = {c\over \gamma W\left({c C^*\over \gamma}{k_BT\over fa}\right)}{k_BT\over f},
\end{eqnarray}
where $C^* = C^{-1/\gamma}$ and $W(z)$, the Lambert Omega function\cite{abramowitz+stegun}, is the inverse of $z = W e^W$. The equation for the FEC for polyelectrolytes becomes
\begin{eqnarray}\label{eq:stretching_equation_full}
x \approx 1 - {\xi_t\over \lambda_K} = 1 - {\gamma\over c}W\left({c C^*\over \gamma}{k_BT\over fa}\right),
\end{eqnarray}
or equivalently,
\begin{eqnarray}\label{eq:FEC_Elect}
f \approx {C^* e^{-{c/\gamma}}}{k_BT\over a}{\exp{\left({c\over\gamma}x\right)}\over 1-x}.
\end{eqnarray}
Thus, in the regime where $x$ is not too close to 1, the $\exp\left({c\over\gamma}x\right)$ factor dominates in Eq.~\eqref{eq:FEC_Elect}, which implies that $f\sim \exp(x)$, or equivalently $x\sim \ln f$. The condition for observing the unusual behavior is $c/\gamma \gg 1$. We show using simulations (see below) that when $1/\gamma \sim \left(\lambda_D/a\right)^{0.36}$, this condition is satisfied for $\lambda_D \gg a$ or at low ionic strengths, just as observed in experiments~\cite{ssDNAPincus,dessinges,Koen}.

\def\bv{{\bold b}}

{\bf Simulations}:\\
In order to validate the theoretical predictions we first performed Monte Carlo (MC) simulations of a PE chain with bending rigidity and screened electrostatic repulsions between the monomers, model that is appropriate for ssDNA. The chain Hamiltonian is given by
\begin{eqnarray}\label{eq:Hamiltonian}
{\mathcal{H}_c\over k_BT} = -{\xi_p^0\over a} \sum_{i=0}^{N} {\bv_i\bv_{i+1}\over a^2} + {q^2\over 2} \lambda_B \!\!\! \mathop{\sum_{i=0}^{N}}_{j=i+1}\!\!\!{e^{-{r_{ij}\over\lambda_D}}\over r_{ij}} - {\bold{f}\over k_BT}\sum_{i=0}^N\bv_i,
\end{eqnarray}
where $N$ is the number of monomers, $\bv_i$ is the $i$-th bond vector (Fig.~\ref{fig:fig1}a), $r_{ij}$ is the distance between beads $i$ and $j$, $\lambda_B=0.7$ nm is the Bjerrum length~\cite{Kremer97}, and $q=a\exp\left(0.0338+1.36 I^{2/5}\right)$nm$^{-1}$ with $I$ measured in molarity units (M), and $q$ is the effective charge number per monomer in monovalent solutions~\cite{YangZhang}. Here $a = 0.55$ nm and $\xi_p^0 = 0.65$ nm, which are similar to the parameters for ssDNA/RNA~\cite{plengthssdna,dessinges,Koen, ssDNAPincus,PhysRevLett.95.268303,ToanMacro2010}. The Debye length is related to $I$ (in M) as $\lambda_D = 1/\sqrt{8\pi\lambda_B 0.602 I}$ in monovalent solutions~\cite{YangZhang}.

Sampling of chain conformations is done using the Metropolis scheme with the crankshaft and pivot moves~\cite{YangZhang,BiophysJ2005}. Starting from an initial conformation, the chain is equilibrated for at least $10^6$ MC steps before the actual sampling starts. The long range nature of the electrostatic interactions, even though dampened by the Debye-Huckel factor, is computationally demanding. To  circumvent this problem, we use a matrix to store all the pair-wise interactions and only update the pairs whose distances are changed by the MC moves. To expedite the computations we introduce a cut-off distance of $20\lambda_D$, which is large enough that it does not introduce any errors. We checked the accuracy by comparing FECs and the tangent correlation data at $I=1$ and 20mM with cut-offs of 10 and 20$\lambda_D$ and no cut-off, and found virtually no discrepancies. With these two enhancements the sampling can be sped up by a factor of at least 5 for large $N$.

\begin{figure}
   \centering
   \includegraphics[width = 5in]{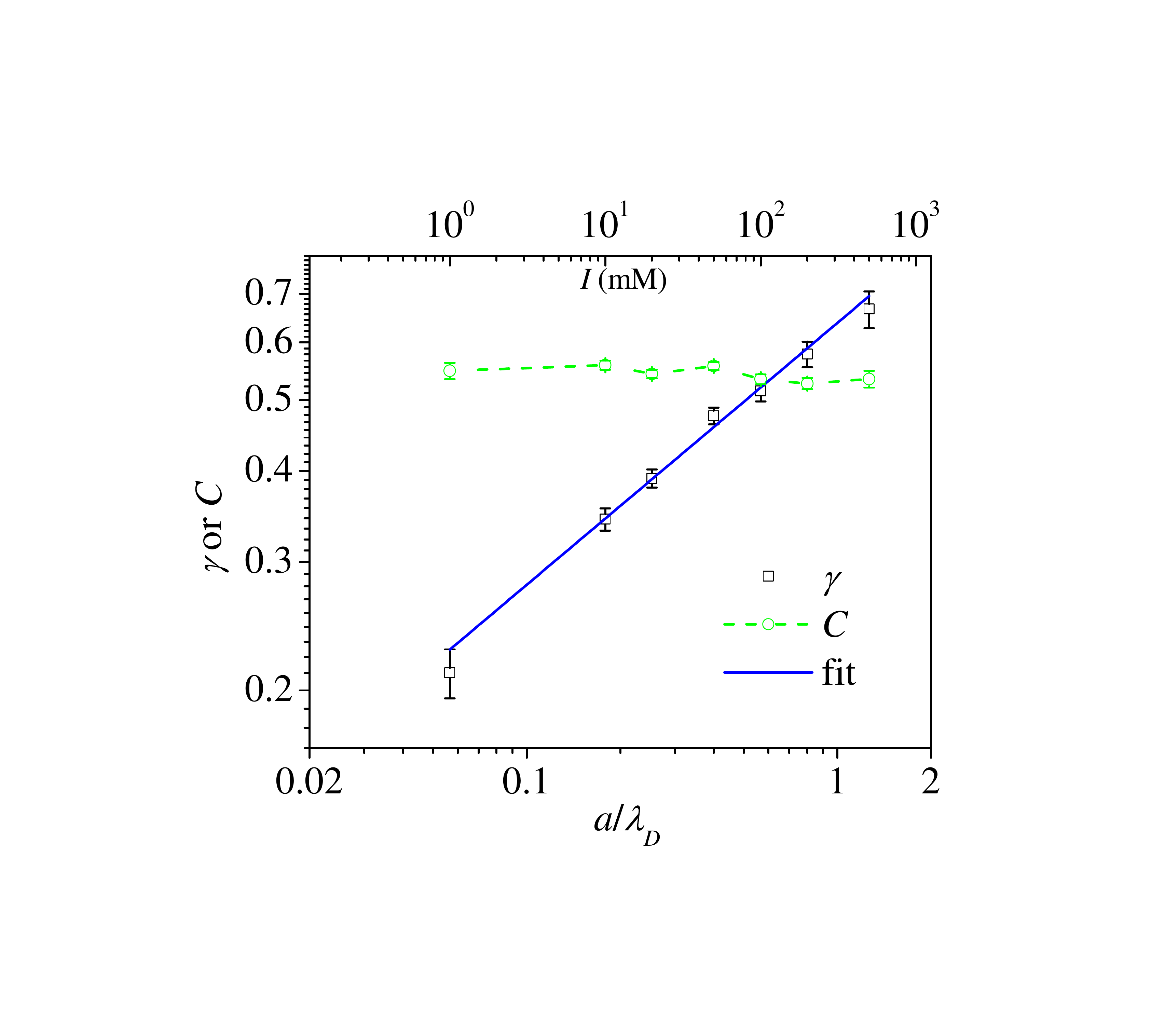}
   \caption{}\label{fig:fig2}
\end{figure}

We calculated the tangent correlation for the model polyelectrolyte for $I$ ranging from $1$ to $500$mM. To reduce finite size effects, we simulated for each $I$ several replicas of the chain with different lengths $N$, starting from 100 and doubling it every time, until the results for $\langle cos(\theta(s) \rangle$ are independent of chain length ($N\ge 800$) (see Fig.~\ref{fig:fig1}b for data at $I=1$mM). The smaller $I$ is  to   (the longer $\lambda_D$) the stronger are the finite size effects. Hence, at low ionic strengths $N$ has to be sufficiently large to obtain converged results. From  Fig.~\ref{fig:fig1}c, it is clear that $\la \cos\theta(s)\ra$ decays exponentially at large $s$ as commonly accepted~\cite{Tricot84,Dobrynin05,Kremer97,Dobrynin09_1}. However, the curvatures at small values of $s$ (especially as salt concentration decreases) suggests that the tangent correlation function qualitatively changes for PEs. The data in Fig.~\ref{fig:fig1}d in a log-log scale clearly shows the postulated power law (Eq.~\eqref{eq:intrinsic_correlation_PE}) at all values of $I$. The range over which the power law decay is observed is from $s/a=1$ up to a few $\lambda_D$'s, which is more robust than we conservatively anticipated. Note that the fit using Eq. (6a) in ref.~\cite{Dobrynin09_1} with a double exponential form, although good for large $s$ (Fig.~\ref{fig:fig1}c, solid line), does not reproduce the true behavior in this regime (Fig.~\ref{fig:fig1}d, solid line). Thus, the simulations validate Eq.~\eqref{eq:intrinsic_correlation_PE}. We fit the initial power law decay to extract $C$ and $\gamma$ as functions of $I$ (or $\lambda_D$). As shown in Fig.~\ref{fig:fig2}, $\gamma \approx 0.64 \left(a/\lambda_D\right)^{0.36}$, whereas $C$ is almost a constant for the range of $I$ considered here.

\begin{figure}
   \centering
   \includegraphics[width = 5in]{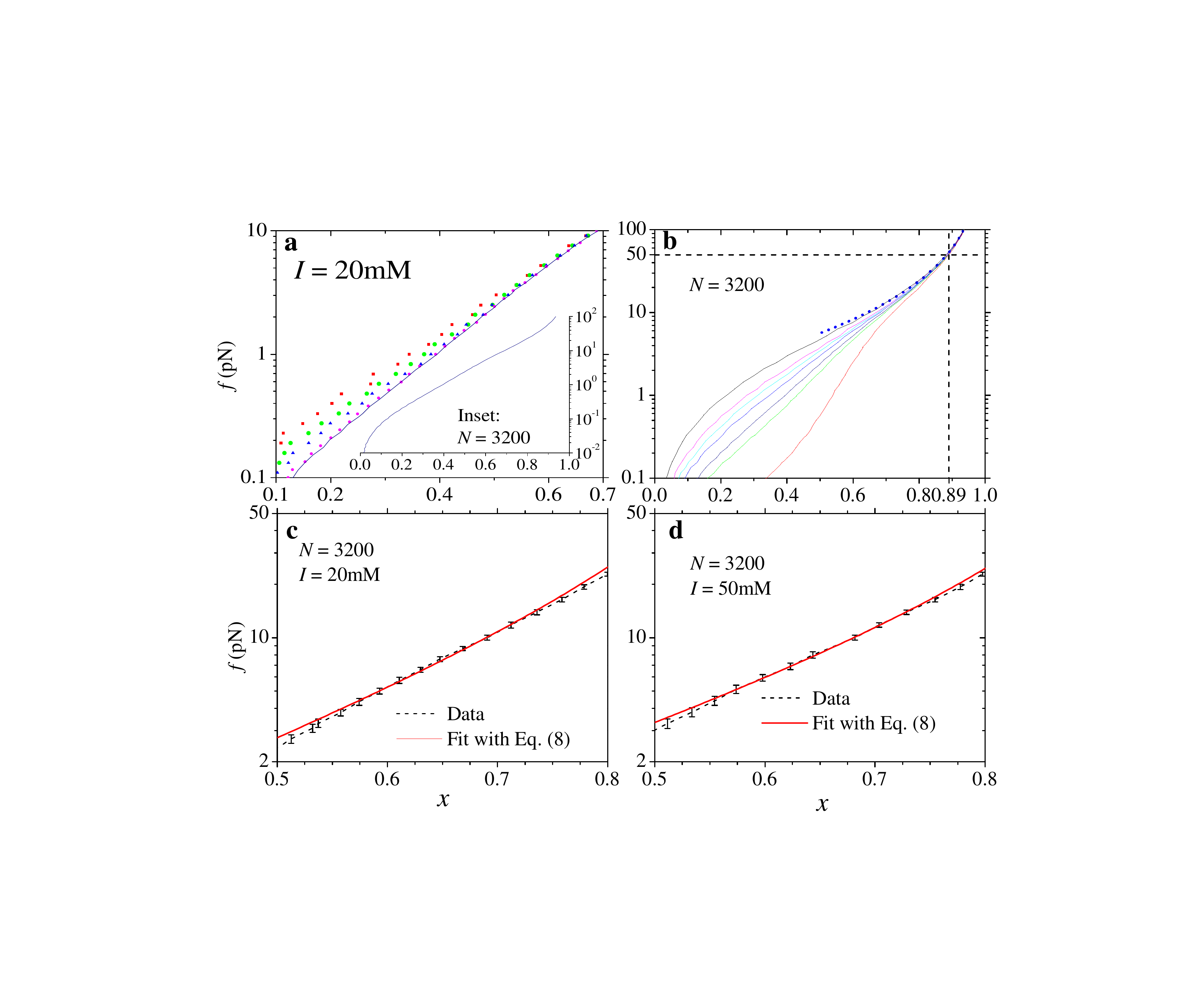}
   \caption{}\label{fig:fig3}
\end{figure}

In order to ascertain the range over which the $x\sim\ln f$ behavior is observed using our model PE, we performed simulations of a chain under tension using $I = 20$mM and $N = 100, 200, \dots, 3200$. As shown in Fig.~\ref{fig:fig3}a, the FECs are almost independent of chain length at $f \gtrsim 5$pN but are clearly $N$-dependent for $f < 5$pN. At even smaller forces, i.e. $f< 0.5$pN, the discrepancies between the $N=1600$ and $N=3200$ curves are still discernible, which imply strong finite size effects. Nevertheless, the prominent feature of the curves is the $x\sim\ln f$ behavior which spans almost 3 decades in force ($0.1$pN$\le f\le 50$pN) for all values of $N$, which becomes increasingly transparent as $N$ increases. Fig.~\ref{fig:fig3}b shows the FECs of a chain with $N=3200$ at several ionic strengths from 1mM to 3000mM. For reference we also show the analytic fit\cite{ToanMacro2010}
\begin{eqnarray}\label{eq:ToanMacro}
x \approx 1 - \xi_t/\sqrt{\lambda_K^2+a^2}
\end{eqnarray}
with $\alpha = 2$ and $c=4$ in the $x>1/2$. It can be seen that the $x\sim\ln f$ regime is visible in most of the curves although it is narrower with increasing ionic strength. Moreover, all the curves to converge at $f\approx 50pN$, or $x \approx 0.89$, which can be analytically fit using Eq.~\eqref{eq:ToanMacro}.
 This implies that all the chains start to enter the FJC regime and electrostatic interactions among the monomers and the ionic conditions play an insignificant role in the high stretching limit. Thus, the observed unusual FEC ($x\sim \ln f$) is pronounced over a few decades in force $f$, and is surprisingly found to be a consequence of the previously unnoticed power law decay of $\la\cos\theta(s)\ra$ (Eq.~4) at short distances.

{\bf Analysis of Experimental Data}

Next we tested the validity of Eq.\eqref{eq:FEC_Elect} by fitting it to the simulated FECs with only $C^*$ and $c/\gamma$ as two adjustable parameters. Because of the approximation made in deriving the equation, we restrict the fitting region to be $0.5< x < 0.8$. Indeed, as shown by the solid lines in Fig.~\ref{fig:fig3}c ($I=20$mM) and Fig.~\ref{fig:fig3}d ($I=50$mM), Eq.\eqref{eq:FEC_Elect} fits the data in the region with reasonable accuracy. At $I=20$mM, the fitted value for $c/\gamma$ is $\sim 4.01$ and for $C^*$ is $\sim 1.78$. If we reconcile the fitted $\gamma$ with the value obtained through independent simulation for the tangent correlation at the same $I$ in Fig.~\ref{fig:fig2}, we get $c=1.7$ and $C\approx 0.80$ ($C$ is larger than the direct-fit value of 0.54). Similar values were also obtained at $I=50$mM.

 Now we use the theory and simulation results to analyze the  stretching data of ssDNA by fitting the FECs. Fig.~4 shows the fit of experimental FECs of ssDNA~\cite{ssDNAPincus} at a few representative ionic strengths $I$ using simulated FECs of our PE model, with the contour length {\em $L_c$ as the only adjustable parameter}. The simulation results  quantitatively reproduce the experimental FECs. The fits at other $I$ from 1mM to 3000mM are also excellent. It should be emphasized that the sole adjustable parameter $L_c$, which varies from one FEC to another, and is very close to the values obtained from the rule from simulated curves at various $I$, $x(f=50$pN$)\approx 0.89$ (see Fig.~\ref{fig:fig3}b).

{\bf Discussions}:

It should be noted that the Hamiltonian~\eqref{eq:Hamiltonian} only takes into account the intrinsic bending rigidity of the polymer and the screened electrostatic interactions as in the Debye-Huckel theory. The ssDNA molecules used in experiments~\cite{ssDNAPincus} were specially treated to prevent base-paring interactions, so it is reasonable not to include them in our model. The high quality of the fits using our theory strongly suggests that  long-ranged interactions  predominantly due to PE effects give rise to the $x \sim \ln f$ behavior . It further validates our arguments that the logarithmic dependence of extension on the force is intimately related to the power law decay in the tangent correlation at small arc-length separations in {\it the absence of force}. Thus, we can readily use the values of $C$ and $\gamma$ in Fig.~\ref{fig:fig2}, that have been extracted from direct simulations of the tangent correlation, as predictions characterizing the decay of tangent correlation of ssDNA.

We can also the  use the approximate stretching equation~\eqref{eq:stretching_equation_full} with $c=1.7$ and $L_c$ obtained from the assumption $x(f=50$pN$)\approx 0.89$, and adjust $C$ and $\gamma$ to fit to the experimental FECs. Using this procedure we also obtained good agreement with the values obtained from a direct fit of Eq.~\eqref{eq:intrinsic_correlation_PE} to the simulation data of tangent correlation. The fitted values of $C$ to the experiment data are somewhat higher than those obtained using the direct fit, but it also stays almost a constant for all the range of $I$ considered. Thus, the analytical theory with only $C$ and $\gamma$ as parameters suffices to fit the experimental data over a wide range of salt concentration.

Logarithmic dependence of ssDNA upon stretching has been reported in earlier single molecule experiments~\cite{dessinges}. However, in fitting the experimental FECs, they used a different model, namely the extensible FJC with electrostatic interactions that are similar to ours in Eq.~\eqref{eq:Hamiltonian}. As noted here, they also showed that the {\em entropic} worm-like chain model~\cite{wlc3} failed to fit the experiment data and tended to give unphysical persistence length. However, as pointed out elsewhere~\cite{ToanMacro2010}, both the discreteness (which is not present in the WLC model, but is in our model and the one by Dessinges et al.~\cite{dessinges}) and the bending stiffness play important roles in the elastic properties of biomolecules in usual ranges of stretching force. Moreover, we have seen that at low ionic strengths, the PE effects in ssDNA are very strong in the range of force less than about 10pN ($I$-dependent). In this range of forces the nature of FECs is  mainly determined by  the electrostatic interactions. At the force higher than 10pN, the semiflexibility of the chain should play an important role and the WLC model~\cite{WLC1,BYHa1997} can be used to quantitatively describe the FECs.

Finally, it is interesting to wonder why the $x\sim \ln f$ dependence has not been observed in dsDNA, which is also an excellent model polyelectrolyte. In dsDNA, the intrinsic persistence length ($\sim$50nm in monovalent salts) is usually much larger than the Debye length, $\lambda_D$. Hence, the range over which the initial power law decay of the tangent correlation can be observed is far too small. For all practical purposes, $\la\cos\theta(s)\ra$ exhibits the standard exponential decay for all $s$. As a result, dsDNA behaves as a WLC with polyelectrolyte effects playing only a minor role especially when the salt concentration is high~\cite{WLC1,thiru}. Thus, the $x\sim \ln f$ behavior cannot be observed in practice in monovalent solutions even though in theory such a regime must exist. In a recent paper, ssDNA in divalent cations such as Mg$^{2+}$ or Ca$^{2+}$ shows even a stronger dependence of extension $x$ on $f$ than $x\sim \ln f$\cite{McIntosh11Macro}. In order to address these issues, we need to develop a better description of ion-PE interactions.

{\bf Conclusions}:

In summary, we have shown that the unusual stretching behavior of ssDNA is due to polyelectrolyte effects whose origins can be traced to a previously unnoticed power law decay of the tangent correlation function in PEs in the absence of force on length scales comparable to the Debye length. Using theoretical arguments and simulations of a minimal model for a PE we demonstrated that at forces (10-50 pN) the FEC is given by $x\sim \ln f$. Our theory quantitatively explains the measured FEC~\cite{ssDNAPincus} ($x\sim \ln f$). The simple analytic expression for FECs can be used to predict the response of force in other single stranded RNA and DNA as well as synthetic polyelectrolytes.

\bigskip

{\bf Acknowledgements}: We are grateful to the National Institutes of Health (GM 089685)for supporting this research. We thank Omar Saleh for discussions and for providing us with his experimental data and a preprint prior to publication.

{\bf Note Added} While this paper was under review we received a preprint by M. J. Stevens, D. B. McIntosh, and O. Saleh  reporting simulations that illustrate the relevance of crumpled structures at short length scales in polyelectrolytes in leading to the observed scaling $x\sim\ln f$. The physics underlying the origin of  $x\sim\ln f$ in both the treatments are similar.
\newpage

\newpage
\clearpage

\begin{figure}
   \centering
   \includegraphics[width = 5in]{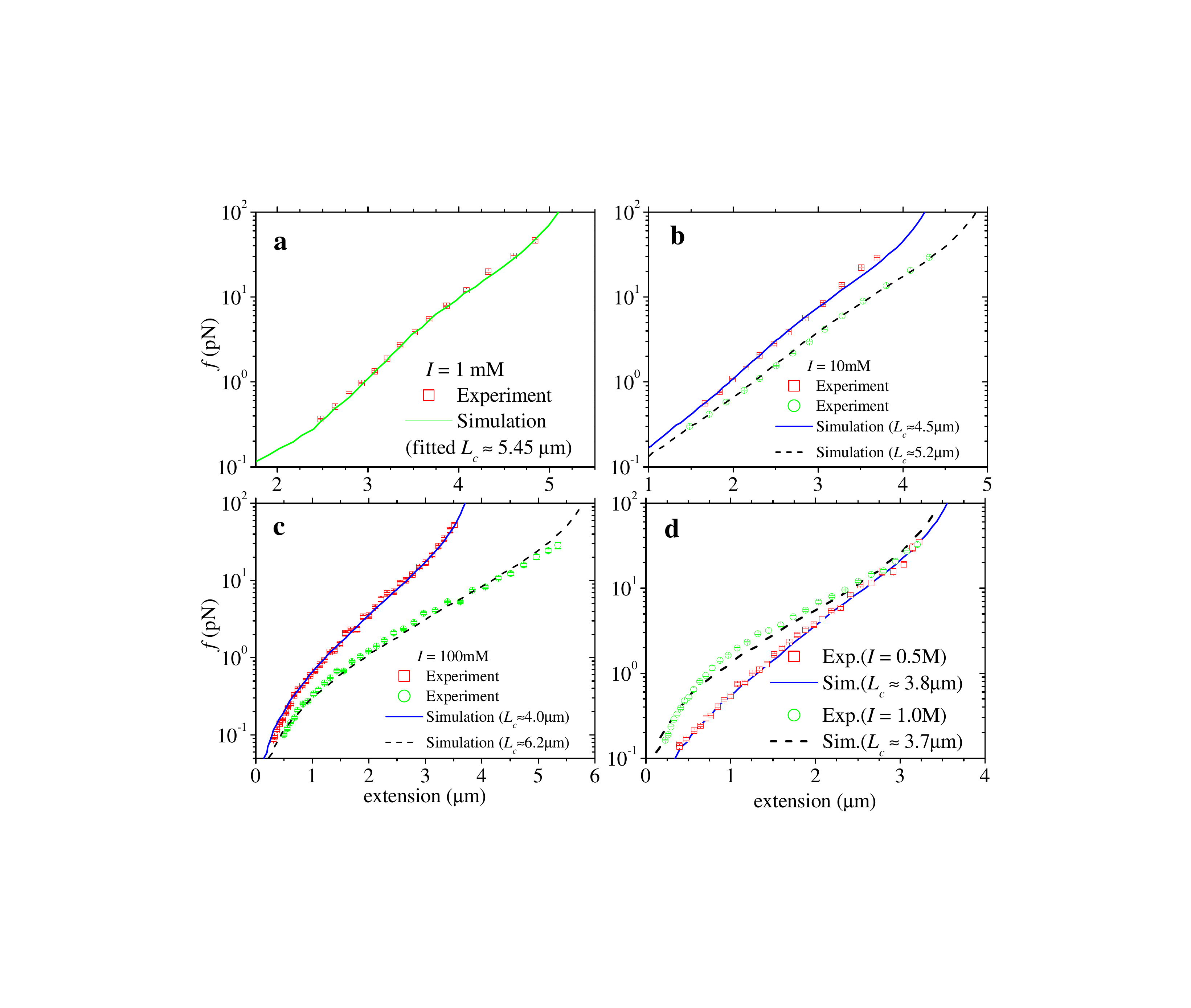}
   \caption{}\label{fig:fig4}
\end{figure}

\begin{figure}
   \centering
   \includegraphics[width = 3in]{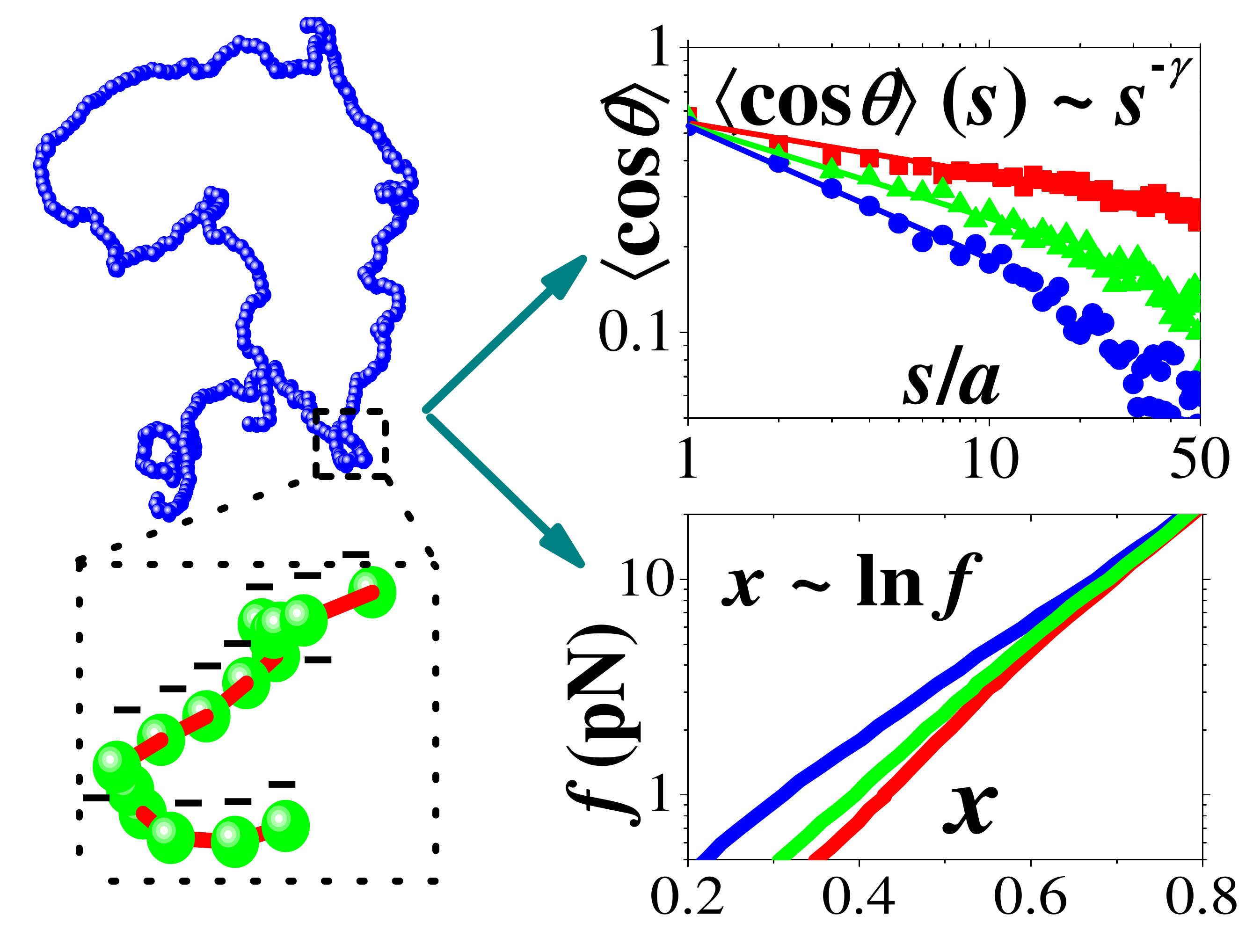}
   \caption{}\nonumber
\end{figure}

\pagebreak
\clearpage
\newpage
{\bf Figure captions:}\\

{\bf Figure~1}: (a) A cartoon of the polymer chain showing tangent vectors at $s_i$ and $s_j$ along the chain. The angle $\theta(s = |s_j-s_i|)$ between the two vectors is shown. The average value over all possible conformation of the chain $\la \cos\theta(s)\ra$ is the tangent correlation along the chain. The length scale over which $\la \cos\theta(s)\ra$ decays at $f=0$ is measure of the chain persistence length. (b) At $I=1$mM, all the curves with various $N$ values (from bottom to top $N$ = 100, 200, 800, 1600) show that there are strong finite size effects. Only when $N$ exceeds 1600 do the curves converge. (c) $\la\cos\theta(s)\ra$ in log-linear scale for $N=1600$ and at various $I$. The solid line is a double exponential fit using $\la \cos\theta\ra = (1-\beta)\exp\left(-s/\lambda_1\right)+\beta\exp\left(-s/\lambda_2\right)$~\cite{Dobrynin09_1}, with $\beta \approx 0.358$ and $\lambda_1 \approx 1.02a$ and $\lambda_2 \approx 179.11a$. (d) To identify the power law behavior the data in (c) is shown as a log-log plot. The curves at various ionic strength clearly show that $\la \cos\theta(s)\ra$ decays as a power law on scale $s \le \lambda_D$. The arrows correspond to $\lambda_D$. The solid line is the double exponential fit.

{\bf Figure~2}: The values of $C$ and $\gamma$ extracted from the simulation data of $\la \cos\theta(s)\ra$ for PE chains with $N=1600$ using Eq.~\eqref{eq:intrinsic_correlation_PE}.

{\bf Figure~3}: (a) Force-extension curves (FECs) for various $N$ values at $I=20$mM. The $N$ values from top to bottom are 100, 200, 400, 1600 and 3200 (solid line). The inset shows the FEC for $N=3200$. (b) FECs for a PE chain with $N=3200$ at various values of $I$. The values of $I$ from bottom to top are 1, 10, 20, 50, 100, 200 and 3000mM.
The dashed line is plotted using $x \approx 1 - \xi_t/\sqrt{\lambda_K^2+a^2}$ Eq.~\eqref{eq:ToanMacro}. (c) and (d): Fits using Eq.~\eqref{eq:FEC_Elect} to simulation data for a chain with $N=3200$ at $I=20$ and 50mM, respectively. See text for the values of the parameters.

{\bf Figure~4}: Single-parameter (using the contour length $L_c$ as the fit parameter) fits of simulated FECs to experimental data of ssDNA in Na$^+$ solutions~\cite{ssDNAPincus}. The high quality of the fits for $I=1$mM (a), 10mM (b), 100mM (c), 500mM and 1000mM (d) indicates that electrostatic interactions play a dominant role in the stretching of ssDNA.


\end{document}